# Seebeck effect in Fe$_{1+x}$Te$_{1-y}$Se$_y$ single crystals

I.Pallecchi[1], G.Lamura[1], M.Tropeano[1], M.Putti[1]
[1] *CNR-INFM-LAMIA and Università di Genova, via Dodecaneso 33, 16146 Genova, Italy*

R.Viennois[2], E.Giannini[2], D.Van der Marel[2]
[2] *DPMC, University of Geneva, 24 quai E.-Ansermet 1211 Geneva, Switzerland*

**Abstract**

We present measurements of resistivity and thermopower S of Fe$_{1+x}$Te$_{1-y}$Se$_y$ single crystalline samples with y=0, 0.1, 0.2, 0.3 and 0.45 in zero field and in a magnetic field B=8T. We find that the shape of thermopower curves appears quite peculiar in respect to that measured in other Fe-based superconducting families. We propose a qualitative analysis of the temperature behavior of S, where the samples are described as almost compensated semimetals: different electron and hole bands with similar carrier concentrations compete and their relative contribution to the thermoelectric transport depends on the respective filling, mobility and coupling with phonons. For y≥0.2, superconductivity occurs and the optimum Se-doping level for a maximum T$_c$ of 13 K turns out to be y=0.3. At low temperatures, evidence of a contribution to S by an excitation-drag mechanism is found, while at high temperatures a strikingly flat behavior of S is explained within a narrow band Hubbard model.

**Introduction**

The newly discovered class of Fe based superconductors [1] is attracting worldwide attention, due to their applicative potential, with transition temperatures above 50 K [2] and very high upper critical fields [3,4], as well as for fundamental studies of superconducting mechanisms, where the role of Fe is yet to be clarified. Various phases have been synthesized to date; all of them share some common features such as square planar sheets of tetrahedrally coordinated Fe ions in their crystal structure and the similar shape of their Fermi surfaces. In particular, in the so called "11" phase of iron chalcogenides (FeCh, Ch=chalcogenide), the Fermi surface is composed by cylindrical electron sections at the zone corners, cylindrical heavy-hole sections around the zone center and, depending on the stoichiometry, other small hole pockets at the zone center [5]. Such topology fulfils the condition for nesting and is thought to give rise to a spin-density-wave (SDW) ground state. The nesting condition is gradually lost upon doping of holes or electrons. The "11" phase is one of the most studied, due to its simple structure, the possibility of growing fairly large single crystals [6] and the reduced toxicity of its constituents compared to As. However, this phase exhibits peculiar properties as compared to other Fe-based phases, namely the non-collinear orientation of anti-ferromagnetic ordering vector and nesting vector [7,8] and no clear signatures in favor of a SDW gap [9,10,12]. Concerning the former point, it has been suggested either that magnetism and superconductivity arise from different interactions or that Cooper pairing does not originate from the itinerant spin fluctuations that accompany the suppression of the SDW state [10], as was instead initially suggested [5,11], in analogy with the other Fe-based phases. Concerning the gapless SDW state, it has been suggested that the ground state of Fe chalcogenides is that of a nearly electron-hole compensated semimetal [10], consistently with Hall effect data [12].

Excess Fe occupying randomly octahedral sites seems to be ever present in synthesized compounds of the 11 phase [12] and it has a dramatic effect on the ground state of the compound [7,8]. Excess Fe is thought to be in the Fe$^+$ valence state so that it dopes the system with electrons; moreover, it has a large magnetic moment (2.4 $\mu_B$) which may have a pair-breaking effect [11].

The 11 compounds undergo a structural and magnetic transition at the same temperature, which is around 65K and is insensitive to the application of a magnetic field [12]. Upon doping, for example with Se or S substitution, superconductivity occurs [6,13,14].

In order to shed light on the peculiar character of the Fe chalcogenides as compared to other Fe-based superconductors, measurements of the Seebeck coefficient S may offer precious information about charge carriers, Fermi surface, density of states and scattering mechanisms in non-superconducting parent compounds as well as in the normal state of doped compounds, where the isovalent substitution of Te by Se brings about the superconductivity. Such information would be complementary to that provided by Hall effect, resistivity and magnetic measurements. Yet, to the best of our knowledge, only one Seebeck effect measurement has been reported on a pure FeSe polycrystalline sample [15], but no measurements on the Fe(Te,Se) system, to date. Even a set of Hall effect data of a series of Fe(Te,Se) samples with different Se content is still missing in literature. On the other hand, several measurements of Seebeck effect on other Fe-based phases have been presented [16,17,18,19,20,21,22,23,24]. Common features are a maximum of the magnitude of the Seebeck coefficient S in proximity of the structural/magnetic transition, generally attributed to a sharp change in the scattering mechanism, as well as a change of sign of S, attributed to the multiband character of transport. The fact that the 11 phase presents no SDW gap and thereby no carrier condensation below the transition may be in principle responsible of a significantly different behavior of S, as compared to other Fe-based phases. In the present work we present resistivity and Seebeck effect measurements on single crystals of various compositions $Fe_{1+x}Te_{1-y}Se_y$.

**Experimental**

We prepared $Fe_{1+x}Te_{1-y}Se_y$ single crystalline samples by the Bridgeman method. Pieces of highly pure Fe, Te and Se were mixed in nominal ratio Fe:Te:Se=(1-x):(1+y):y (with x=0-0.1 and y = 0, 0.1, 0.2, 0.3, 0.45), then heated to 930-960°C in sealed quartz reactors and slowly cooled to room temperature. Single-phase, single crystalline samples were obtained for the whole series as proven by X-rays diffraction and SEM/EDX (scanning electron microscopy/energy dispersive X-ray microanalisys) analyses reported elsewhere [25,26]. It is possible that our transport measurements probe a percolative path of minimum resistivity, more likely at the crystal surface, whose local properties depart from those of the bulk sample. This concerns mainly the local excess Fe, as compared to the average excess Fe. The actual average excess Fe is not the same for all the samples and was estimated by the Rietveld refinement of X-rays diffraction data, as reported in Table I. It ranges from 0.017 to 0.087. We have found that only in the case of y=0.3, the excess Fe content has a well visible effect in the transport properties, such as resistivity and Seebeck effect, whereas in the other cases transport properties are determined mainly by the Se content and only weakly affected by the Fe stoichiometry. Nevertheless, it is worth remembering that even small changes in the actual Fe content can have a non negligible influence on some properties, such as the magnetic susceptibility, as reported elsewhere [25,26].

We measured both resistivity and Seebeck effect in a Quantum Design PPMS system as a function of temperature from 5K to 300K. In the case of the S measurement, we applied a square-wave heat flow along the Fe planes, with adjustable period (from 400 s to 1450 s) and thermal gradient (from 0.1K to few K), and measured the related voltage drop decay; the measurements were carried out both in zero field and in a magnetic field of 8T, parallel to the Fe planes.

| Nominal composition | Refined composition |
|---|---|
| FeTe | $Fe_{1.087}Te$ |
| $FeTe_{0.8}Se_{0.2}$ | $Fe_{1.049}Te_{0.8}Se_{0.2}$ |
| $FeTe_{0.7}Se_{0.3}$ (a) | $Fe_{1.053}Te_{0.7}Se_{0.3}$ |
| $Fe_{0.9}Te_{0.7}Se_{0.3}$ (b) | $Fe_{1.013}Te_{0.7}Se_{0.3}$ |

**Table I:** Values of excess Fe content as compared to nominal compositions, obtained by Rietveld refinement of X-rays data in samples nominally identical to the ones measured in this work. Only four out of six samples presented in this work are listed, because in two cases Rietveld refinement could not be carried out, due to the lower structural quality.

## Results

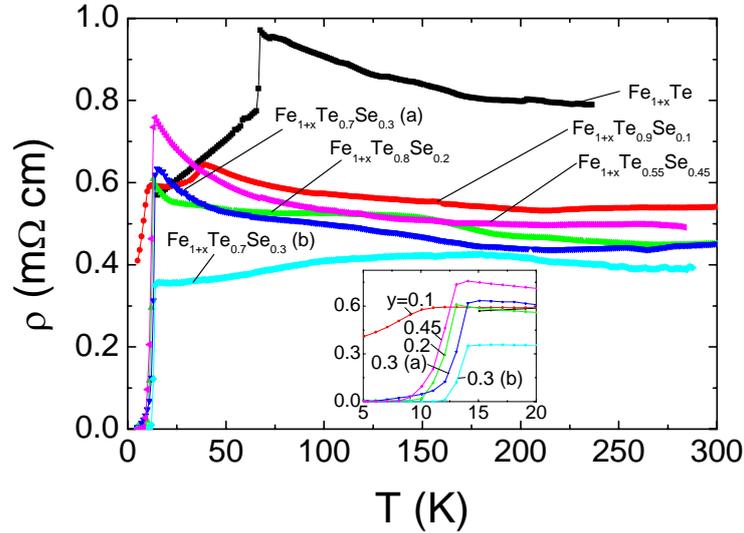

**Figure 1:** (color online) Resistivity as a function of temperature for the six samples; inset: zoom of the low temperature region.

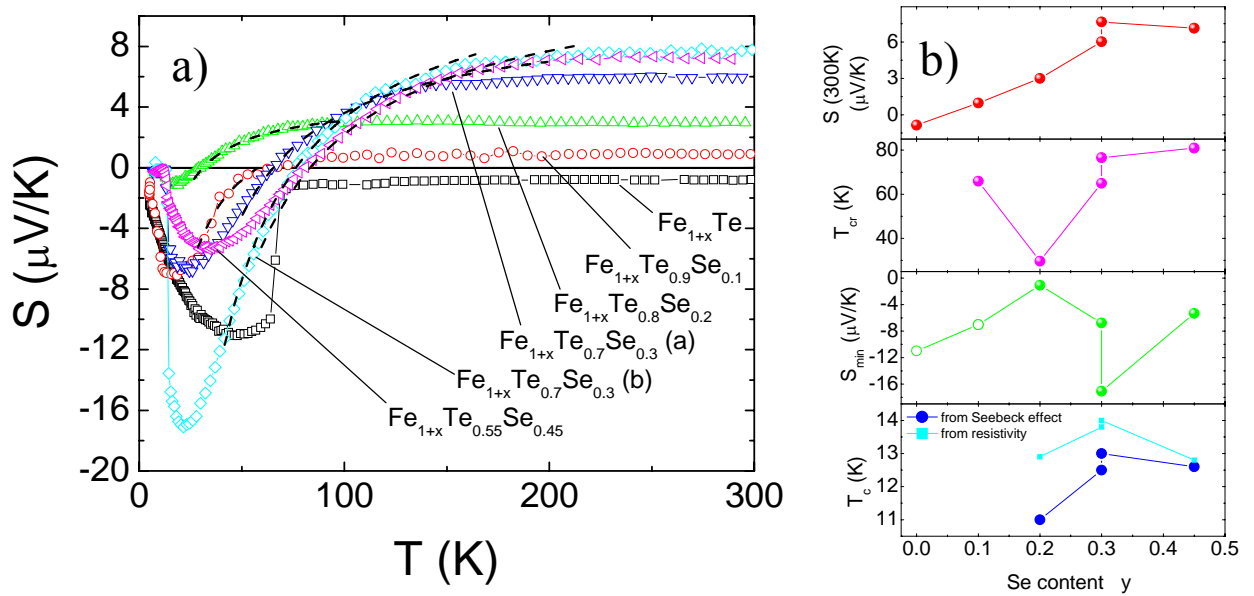

**Figure 2:** (color online) a) Seebeck coefficient as a function of temperature for the six samples; dashed black lines show the fit $S=A+B/T$, with A and B fitting constants. b) From top to bottom, the following quantities are plotted as a function of the Se content y (atoms per unit formula): value of the Seebeck coefficient at T=300K, crossover temperature at which S changes in sign, value of the Seebeck coefficient at the minimum of the S curve and superconducting transition temperature, extracted from resistivity (90% of normal state resistivity criterion) and Seebeck measurements. In the case of $S_{min}$, different symbols have been used for superconducting and non superconducting samples, in order to emphasize visually the correspondence of $S_{min}$ and $T_c$ in superconducting samples, described in the text.

In figure 1, the resistivities of the six samples are presented. The $Fe_{1+x}Te$ sample presents a discontinuity and an abrupt change of slope at the magnetic/structural transition at 66K. In the $Fe_{1+x}Te_{0.9}Se_{0.1}$ sample, a similar feature, shifted at a lower temperature ~39K, is seen. Moreover, the onset of a superconducting transition appears at 10K, even if the resistivity does not vanish above 5K. All the other samples present a superconducting transition with vanishing resistivity. The highest transition temperature $T_c$ is found in the $Fe_{1+x}Te_{0.7}Se_{0.3}$ samples, as shown in the bottom panel of figure 2b). In particular, of the two different samples with y=0.3 and bulk excess Fe content x=0.053 and x=0.013, called sample (a) and (b) respectively, the optimal one in terms of $T_c$ turns out to be the latter. This optimal sample is also characterized by a metallic slope as a function of temperature above $T_c$, whereas all the other samples exhibit a weak negative slope of resistivity as a function of temperature. This result is consistent with magnetic measurements on the same samples [25,26], which show bulk superconductivity only for this optimal composition. We notice that bulk superconductivity has been found in a $Fe_{1+x}Te_{0.6}Se_{0.4}$ sample in ref. [27], whereas no sign of bulk superconductivity has been found by specific heat measurements in a $Fe_{1+x}Te_{0.55}Se_{0.45}$ sample, in ref. [6]; indeed, this seems to be the limit Se content below which bulk superconductivity may or may not appear, depending on the excess Fe content.

In figure 2a) we present thermopower curves of the six samples as a function of temperature. It can be seen that in the high temperature regime, differently from all the other Fe-based superconductors, the Seebeck coefficient S is nearly constant; its value increases monotonically with increasing Se content y, from a negative value S(300K)=-0.85μV/K for the y=0 sample to a positive saturation value slightly above +7μV/K for the samples with y>0.3. This trend is also plotted in the uppermost panel of figure 2b).

In the $Fe_{1+x}Te$ sample, S undergoes an abrupt step-like change at the magnetic/structural transition around 66K, in agreement with the step in the resistivity curve, it reaches a minimum value and eventually tends to vanish when the temperature tends to zero. The y=0.1 sample is reminiscent of the same behavior, but the transition appears to be broadened and the transition temperature suppressed below 50K. The curve passes from a positive constant value around +0.97μV/K to a minimum negative value; the crossover temperature $T_{cr}$ where S changes in sign is 66K. The other samples with y≥0.2 follow the same trend described above in the high temperature regime, but are pretty different at the lowest temperatures: consistent with resistivity curves, the Seebeck coefficient becomes zero at finite temperatures, indicating the onset of the superconducting state. The temperature $T_c$ where S vanishes is around 11K for the y=0.2 sample, it reaches 13K with increasing y and then slightly decreases in the y=0.45 sample. This trend is plotted in the bottom panel of figure 2b) and confirms that y=0.3 is the optimal Se concentration for superconductivity. The transition temperatures extracted from resistivity and Seebeck effect are slightly different, either for reasons related to the measuring technique, as the Seebeck measurement is much slower and averages out the signal over a larger temperature interval during the temperature sweep, or due to the particular criterion chosen to define $T_c$, or else due to intrinsic reasons related to the difference in these properties, as noted in ref. 24. However the trend of $T_c$ as a function of y is alike. Above the superconducting transition, the samples with 0.2≤y≤0.45 exhibit a similar behavior, with a negative minimum of S, a crossover at $T_{cr}$ where S changes in sign, and a constant positive high temperature value. The crossover temperature increases monotonically with increasing y≥0.2, as shown in the second panel of figure 2b), while the minimum value of S increases in magnitude with y for the 0.2≤y<0.45 samples, but it decreases for the y=0.45 sample, as shown in the third panel of figure 2b).

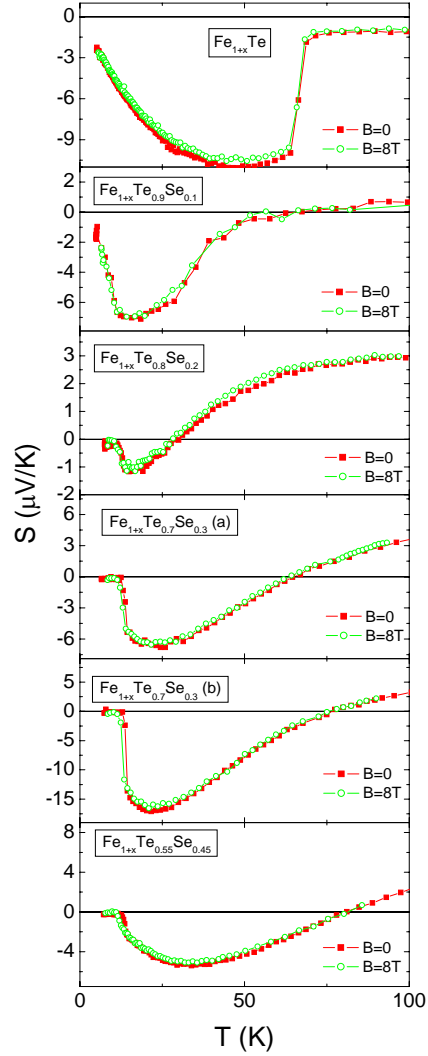

**Figure 3:** (color online) Seebeck coefficient as a function of temperature for the six samples in the low temperature region, in zero field and in a magnetic field B=8T.

The effect of the application of a magnetic field is very weak in all cases. In figure 3, the S curve at B=0 and B=8T are compared for all the six samples. In the high temperature regime, not shown, the effect of the magnetic field is negligible and the B=0 and B=8T curves merge within the experimental uncertainty. In the low temperature regime around the negative minimum of S, the B=8T curves are systematically smaller in magnitude than the B=0 curves. Finally, the 8T field slightly lowers the superconducting transition temperature; the shift of $T_c$ is smaller than 1K for all samples, consistently with the very large upper critical field $H_{c2}^{\parallel}$ parallel to the Fe planes reported in literature [12,3].

**Discussion**

The Seebeck coefficient in a metal in the diffusive regime can be expressed as [28]:

$$S = \frac{\pi^2 K_B^2 T}{3q} \left( \frac{1}{\sigma(E)} \frac{d\sigma(E)}{dE} \right)\bigg|_{E=E_F} \approx \frac{\pi^2 K_B^2 T}{3q} \left( \frac{N}{n} + \frac{1}{\tau} \frac{d\tau}{dE} \right)\bigg|_{E=E_F} \quad (1)$$

where $K_B$ is the Boltzmann constant, N is the density of states, $E_F$ is the Fermi energy, $q$ is the electron charge (with positive sign for hole transport and negative sign for electron transport), $n$ is the charge carrier density, $\sigma$ is the conductivity and $\tau$ is the scattering time. Usually the first term is dominant, so that S has the same sign of the charge carriers. The second term depends on the scattering mechanism and is negative or negligible, as the energy dependence of τ is usually described by a power law $\tau \propto E^{-\alpha}$ with $\alpha>0$. Yet, close to a transition, if any abrupt change in the scattering mechanism occurs, also the second term may play a role.

When more than one band cross the Fermi level and contribute to the transport, eq. (1) describes the contribution of each band to the Seebeck effect, and the overall Seebeck effect is the sum of the band contributions, weighted by the respective band conductivities:

$$S = \frac{\sum \sigma_i S_i}{\sum \sigma_i} \qquad (2)$$

where $i$ is the band index and $\sigma_i$ and $S_i$ are the electrical conductivity and Seebeck coefficient of the *i-th* band, respectively. In the case of Fe(Te,Se) compounds, four or five bands are indeed thought to contribute to transport, depending on the composition [5,10].

Considering the curves of figure 2a) and specifically the high temperature behavior plotted in the topmost panel of figure 2b), it can be said that the isovalent substitution of Te with Se results in an effect equivalent to introducing holes in the system, so that the high temperature Seebeck coefficient passes from negative to positive values with increasing y. In parallel, the resistivity decreases with increasing y. In fact, the structure is stabilized by either excess Fe or Se substitution; hence, with increasing Se substitution, less excess Fe is required [26]. As excess Fe is in the +1 valence state, it dopes electrons into the system, so that the effect of Se substitution results in less electron doping and is equivalent to hole doping. Beside this scenario, this hole doping could be alternatively explained as an effect of band modification by isovalent substitution of Se. Regarding the effect of hole doping on the Seebeck value at high temperatures, it is clearly understood in terms of eq. (2), by assuming that with increasing y the conductivities of the hole bands increase and the high temperature behavior of the thermopower is eventually dominated by the positive Seebeck coefficients of the hole bands. We notice that in a single band metal, the first term of eq. (1) yields a decrease of S with increasing doping, so that the trend of S(300K) as a function of *y* can be accounted for only within a multiband picture, where the relative contributions of the different bands change with *y* [29]. This effect is particularly dramatic in all Fe-based superconductors, due to their nature of almost compensated metals, with bands of carriers of opposite signs and nearly equal charge concentrations. In the 11 phase, the small magnitude of the Seebeck coefficient, which hardly exceeds 10μV/K, as compared to those of other Fe-based phases, that are several tens of μV/K [16], suggests that this charge compensation is even more balanced.

The strictly constant dependence of S on temperature at high temperatures is remarkable and peculiar with respect to other Fe-based families. Such behavior has been predicted in the narrow band Hubbard model, valid for semiconductors and metals, at sufficiently high temperatures, where the kinetic terms of S can be neglected [30,31]; hence the high temperature flatness of the S curves seems to point out that in iron chalcogenides transport has a more localized character than in other Fe-based families. Such higher localization is also confirmed by magnetic susceptivity data on the same samples, showing a more evident Curie-Weiss behaviour [25,26]. In the case of strong on-site Coulomb repulsion [32] $U \gg K_B T$, the model predicts:

$$S(T \to \infty) = \frac{K_B}{q} \ln\left(2\frac{1-n}{n}\right) \qquad (3)$$

where *n* is the carrier density per atomic site. This description has been applied to account for the high temperature constant behaviour of S in metallic $La_2CuO_{4-\delta}$ [33] and semiconducting $La_2CuO_4$ [34]. This model has been developed only for single band systems, which is not our case; if we try to apply it to our data, from our S values at high temperature, we get that n varies by 3% around 0.67 carriers/site through our series of samples. This value of 0.67 carriers/site must not be interpreted as the actual carrier concentration, because obviously it cannot alone describe the complexity of multiband transport. However, we notice that it is in perfect agreement with the value that is obtained by applying a single band formula $n=1/(qR_H)$ to the Hall resistance $R_H$ data of ref. 12. This consistency suggests the plausibility of the application of the above narrow band Hubbard model to this system, to account for the flat high temperature behaviour of S.

Also the crossover temperature $T_{cr}$ where S changes in sign can be interpreted in terms of a multi-band picture. Below $T_{cr}$ the electron bands dominate the thermoelectrical transport and S is negative; however, if these electron bands are more strongly coupled with phonons or other excitations than the hole bands, their conductivity has a steeper temperature dependence and eventually, above $T_{cr}$, thermoelectrical transport becomes dominated by the hole bands. In this scenario, the increase of $T_{cr}$ with y, shown in the second panel of figure 2b), indicates that, despite the relative weight of holes in the thermoelectrical transport at high temperatures increases with increasing y, the temperature range over which electrons dominate becomes broader. These two findings could be reconciled in the hypothesis that Se substitution is effective in either decreasing the coupling with excitations of carriers in the electron channel or in increasing the coupling in the hole channel. Indeed, within each band, the temperature dependence of scattering rate is proportional to the coupling strength; thereby, if the coupling of hole bands increases with y, the scattering rate in hole bands increases with temperature more quickly for larger y, so that the hole bands start to dominate thermoelectrical transport only beyond a larger value of the $T_{cr}$ temperature. This hypothesis of increasing hole band coupling with increasing y seems consistent with the identification of such coupling with the pairing mechanism responsible for the occurrence of superconductivity in the Se substituted samples.

In the $Fe_{1+x}Te$ sample, the abrupt change of S at the magnetic and structural transition is likely again an indication that the relative contribution to transport of the different bands undergoes a dramatic change. A more or less sharp change of S at the transition is a common feature of all Fe-based superconductors and respective parent compounds [16,17,18,19,20,21,22,23,24]; a non monotonic behavior with local minima and maxima is often observed [16,19,20,23,24], pointing to the complexity of the competing mechanisms into play. Also in our case, other mechanisms, besides this rearrangement of multiband contributions, may yield significant effects, even if the simultaneous occurrence of the structural and magnetic transitions in iron chalcogenides may make it difficult to discriminate the different contributions. Firstly, a sudden change in the scattering mechanism, that is the second term of eq. (1), may undergo an abrupt change at the transition; indeed, scattering by spin fluctuations associated to either the SDW state or the AF ordering may change dramatically. Secondly, the spin-entropy term of S may be affected by the magnetic ordering, which could indeed limit the energetically allowable spin configurations of carriers, due to the exchange interaction between the magnetic ions and the carriers themselves. The change of slope observed in the temperature dependence of resistivity (see $Fe_{1+x}Te$ curve in figure 1) can be reconciled with either the above scenarios of rearrangement of multiband contributions and change in the scattering mechanism. However, the Hall effect measurements reported in literature are more consistent with the band rearrangement picture, as the Hall resistance exhibits an abrupt step at the transition either with [12] or without [13] a change in sign, possibly depending on the excess Fe content.

The sample $Fe_{1+x}Te_{0.9}Se_{0.1}$ can be considered to be midway between the behavior of the parent compound and that of the superconducting samples with y≥0.2. Indeed, for 0.2≤y≤0.3, a monotonic trend of increasing $T_{cr}$ with increasing y is observed. The magnitude of S at the minimum, $|S_{min}|$, and $T_c$ increase with increasing y≥0.2, but they decrease again for the y=0.45 sample. $|S_{min}|$ may be viewed as a measure of the charge density unbalance between the electrons and hole bands; this unbalance is thought to destroy the nesting condition for the formation of the spin-density-wave ground state, thus favoring the onset of the superconducting ground state. Indeed, the sample with the largest $|S_{min}|$ value, that is the $Fe_{1+x}Te_{0.7}Se_{0.3}$ sample (sample b), also has bulk superconductivity, as seen in magnetic measurements [25,26], and the largest $T_c$. Yet we suggest another interpretation of this proportionality between $T_c$ and $|S_{min}|$, namely that at low temperatures, beside the diffusive term of S expressed by eq. (1), there is a phonon-drag term [35], or, more in general, an excitation-drag term. Such a term may indeed reveal the presence of a strong interaction between electrons and excitations, possibly spin fluctuations, which may be identified with the pairing mechanism that ultimately determine the $T_c$. The examination of the shape of the S curves in the range between the temperature of the minimum of S ($T_{min}$) and the saturation temperature allows to substantiate this hypothesis: indeed, the curves of all the samples, except the $Fe_{1+x}Te$ one, are very well described by a power law $\propto T^{-1}$ in the range from $\approx 1.5 T_{min}$ up to the onset of the flat behavior (see dashed black lines in figure 2a)); this law is just a signature of a drag term in the thermopower [36].

The large magnetic field of 8T, besides shifting the superconducting transition temperature proportionally to $(dH_{c2}^{\parallel}/dT_c)^{-1} \approx 0.1$ K/T, has the effect of decreasing the magnitude of S at low temperature by few percent. This dependence is opposite to that observed in SmAsFeO and NdAsFeO oxypnicides [23]. In our case, it may be related to the magnetic field dependence of the spin fluctuation-drag contribution to the thermopower: as the spin fluctuations are quenched by the magnetic field, they drag the electrons less than at B=0. This has been observed experimentally in different systems with spin fluctuations [37,38] and was theoretically predicted by Granemann [39]. Such spin fluctuations coupled with the charge carriers may be identified with the incommensurate two dimensional spin excitations that have been found by inelastic neutron scattering measurements [40]; their contribution dominates the spin fluctuation spectrum and is possibly responsible for the pairing mechanism.

All the above discussions are qualitative and based on conjectures. A theoretical backing could provide valuable information about conductivities, carrier density and phonon couplings of each band for different chemical compositions, which would be necessary ingredients for quantitative analysis of our experimental data. This could help in casting light on the electronic mechanisms which determine the ground state in these 11 phase samples and more in general in all Fe-based phases. In parallel, experimental measurements of other transport properties such as magnetoresistivity and Hall effect on single crystal samples or on epitaxial thin films would help in completing a consistent picture.

**Conclusions**

We measure the Seebeck coefficient curves of $Fe_{1+x}Te_{1-y}Se_y$ single crystalline samples with y=0, 0.1, 0.2, 0.3 and 0.45 in zero field and in a magnetic field B=8T, in order to evidence the peculiarities of the so called "11" phase among the Fe-based superconductors. Indeed, Seebeck curves exhibit a fairly different shape as respect to the ones measured in the other phases. We find that Se doping favors hole transport in the system, however different bands of both electrons and holes contribute to the total Seebeck coefficient. Due to the almost compensated metallic character of these systems, the relative contribution of the different bands yields dramatic changes of S as a function of temperature and doping, yielding a superconducting ground state for y≥0.2. The optimal Se concentration for superconductivity turns out to be y=0.3, with an excess Fe content x=0.013. At low temperatures, an excitation-drag contribution to the thermopower, which is possibly closely related to the excitation-mediated pairing, is identified; its weak but detectable magnetic field dependence suggests that such excitations are spin fluctuations. At high temperatures, the Seebeck effect is strictly flat, which is a peculiar behavior among other Fe-based families and points to a more localized type of transport in iron chalcogenides.

**Acknowledgements**

This work was partially supported by Compagnia di S. Paolo.